\documentclass[aps,prfluids,reprint,onecolumn]{revtex4-2}

\usepackage{amsmath,color,graphicx,amssymb}
\pdfoutput=1
\usepackage{natbib}
\usepackage{amsbsy}
\usepackage{epstopdf}

\definecolor{purple}{rgb}{0.57,0,0.86}
\definecolor{orange}{rgb}{1,0.5,0}

\newcommand{\ea}[1]{{{\color{black} #1}}}
\newcommand{\am}[1]{{{\color{black} #1}}}
\newcommand{\blue}[1]{{{\color{blue} #1}}}
\newcommand{\red}[1]{{{\color{red} #1}}}

\begin{document}

\title{Intermittent Unsteady Propulsion with a Combined Heaving and Pitching Foil}% Force line breaks with \\

\author{Emre Akoz}
 \email{akozemr@gmail.com}

\author{Amin Mivehchi}
 
\author{Keith W. Moored}%
\affiliation{Mechanical Engineering and Mechanics, Lehigh University, Bethlehem, PA, 18015 USA
}

\date{\today}% It is always \today, today,

\begin{abstract}
Inviscid computations are presented of a self-propelled virtual body connected to a combined heaving and pitching foil that uses continuous and intermittent motions.  It is determined that intermittent swimming can improve efficiency when the dimensionless \ea{heave ratio} is $h^* < 0.7$ while it degrades efficiency for $h^* \geq 0.7$.  This is a consequence of the physical origins of the force production for pitch-dominated ($h^* < 0.5$) and heave-dominated ($h^* > 0.5$) motions. Based on insight derived from classic unsteady thin airfoil theory, it is discovered that pitch-dominated motions are driven by added-mass-based thrust production where self-propelled efficiency is maximized for high reduced frequencies, while heave-dominated motions are driven by circulatory-based thrust production where self-propelled efficiency is maximized by low reduced frequencies. Regardless of the dimensionless \ea{heave ratio} the reduced frequency is high for small amplitude motions, high Lighthill numbers, and low duty cycles and \textit{vice versa}.  Moreover, during intermittent swimming, the stopping vortex that is shed at the junction of the bursting and coasting phases becomes negligibly weak for $h^* < 0.5$ and small amplitude motions of $A^* = 0.4$.  This study provides insight into the mechanistic trade-offs that occur when biological or bio-inspired swimmers continuously or intermittently use combined heaving and pitching hydrofoils.  
\end{abstract}

\maketitle

%\tableofcontents

%\section{\label{sec:level1}First-level heading}

%This sample document demonstrates proper use of REV\TeX~4.1 (and \LaTeXe) in mansucripts prepared for submission to APS journals. Further information can be found in the REV\TeX~4.1 documentation included in the distribution or available at \url{http://authors.aps.org/revtex4/}.

%When commands are referred to in this example file, they are always shown with their required arguments, using normal \TeX{} format. In this format, \verb+#1+, \verb+#2+, etc. stand for required author-supplied arguments to commands. For example, in \verb+\section{#1}+ the \verb+#1+ stands for the title text of the author's section heading, and in \verb+\title{#1}+ the \verb+#1+stands for the title text of the paper.

%Line breaks in section headings at all levels can be introduced using \textbackslash\textbackslash. A blank input line tells \TeX\ that the paragraph has ended. Note that top-level section headings are automatically uppercased. If a specific letter or word should appear in lowercase instead, you must escape it using \verb+\lowercase{#1}+ as in the word ``via'' above.

%%-------------------------------------------------INTRODUCTION---------------------------------------------------%%
\section{\label{sec:level1}Introduction}
Many fish and marine mammals use intermittent locomotion, also known as burst-and-coast or burst-and-glide swimming, where they swim with a combination of an active swimming phase and a passive coasting phase \cite{Kramer2001}. The use of intermittent swimming by organisms has been connected to fatigue recovery  during coasting, improved perception, and a decrease in the energetic cost of transport \cite{Videler1982, Kramer2001,  Fish2010}. In fact, theoretical work has shown that intermittent swimming can reduce the energetic cost of swimming by over 50\% when compared to continuous swimming \cite{Weihs1974, Weihs1980}, at least for simplified hydrodynamic models.  This has motivated several experimental studies to directly and indirectly measure the energy savings of intermittent swimming by using live fish \cite{Videler1981, Wu2007, Muller2000}.  Indeed, energy savings of 45\% were observed in live fish, which supported the findings of the previous theoretical work \cite{Wu2007}.  Even numerical modeling of undulating fish found 56\% energy savings when intermittent swimming was employed \cite{Chung2009}. 

Previous work hypothesized that the energy benefit of intermittent swimming was derived from the Bone-Lighthill boundary layer thinning hypothesis \cite{Lighthill1971,Weihs1974}.   This hypothesis supposes that when a fish undulates its body the boundary layers thin due to the wall normal velocity and consequently skin friction drag increases when a fish is actively propelling itself.  Therefore, when fish intersperse passive coasting phases between active undulating phases they can reduce their drag and energy expenditure over a burst-and-coast period.  Indeed, recent work has confirmed the Bone-Lighthill boundary layer thinning hypothesis, however, the drag increase is much more modest on the order of 20-70\% \cite{Ehrenstein2013,Ehrenstein2014} as opposed to original estimates of a factor of 4-10 increase \cite{Lighthill1971, Webb1975, Videler1981, Wu2007}.  With more modest drag increases, this viscous mechanism cannot account for the full energy savings observed in fish and numerical simulations \cite{Akoz2018}, prompting recent work to revisit the basic hydrodynamic mechanisms of intermittent swimming.  \citet{Akoz2018} discovered an inviscid Garrick mechanism responsible for up to 60\% energy savings for a purely pitching hydrofoil swimming intermittently.  They further developed scaling laws to capture the thrust and power performance of intermittently pitching foils by introducing physically motivated nonlinear corrections to classic unsteady linear theory \cite{Garrick1936}.  These scaling laws and numerical performance results were further corroborated by experimental measurements \cite{floryan2017} and direct numerical simulations \cite{akoz2019large} of an intermittently pitching foil.  A purely pitching foil is a reasonable first approximation to the pitch-dominated caudal fin motion of typical intermittent swimmers such as trout \cite{Yanase2015,Yanase2016}, koi carps \cite{Wu2007}, and cod \cite{Videler1981}.  However, their caudal fins are not purely pitching and instead are employing combined heaving and pitching motions.  

Continuously swimming combined heaving and pitching foils have been extensively studied \cite{Anderson1998,Read2003,van2018scaling}.  These kinematics produce high efficiencies up to 72\% at angles of attack of $15^o$ \cite{Read2003}, which is substantially higher than pure pitching kinematics with peak efficiencies around 22-28\% \cite{Boschitsch2014,Mackowski2017,Kurt2018}.  Yet, the performance, wake structures and mechanisms invoked by \textit{intermittently} swimming combined heaving and pitching foils have not been examined.  

In light of these observations, this study extends the study of \citet{Akoz2018} to investigate intermittently swimming combined heaving and pitching foils in an inviscid flow. We address three main questions: How does the proportion of heave and pitch affect the propulsive performance in continuous and intermittent swimming? When is it energetically favorable to choose intermittent over continuous locomotion?  What are the mechanisms behind the observed trends when the dimensionless \ea{heave ratio} is varied?  To answer these questions, we systematically investigate the effect of the dimensionless \ea{heave ratio}, Lighthill number, reduced frequency and amplitude of motion on intermittent swimming.  The paper is organized in the following manner.  Section \ref{s:prob} defines the problem formulation, the numerical modeling approach employed, and the input/output variables and parameters.  Section \ref{s:results} presents the results.  Section \ref{s:concl} summarizes the conclusions of the study.

%%---------------------------------------PROBLEM FORMULATION---------------------------------------------------%%

\section{Problem Formulation} \label{s:prob}
\subsection{Input Variables and Parameters}
\begin{figure}[t]
    \centering
    \includegraphics[width = 0.75\textwidth]{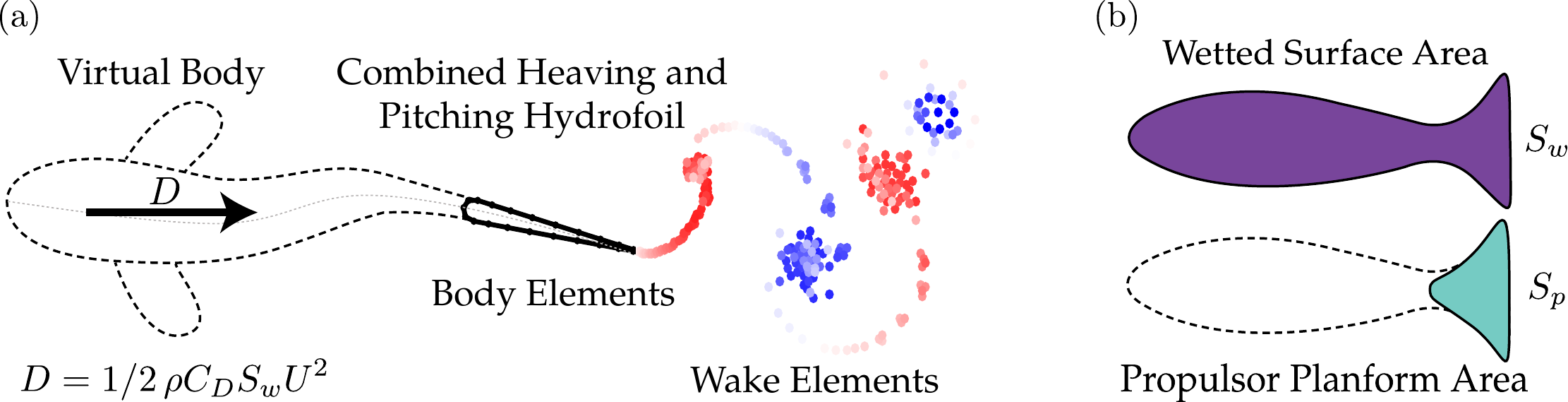}
    \caption{(a) Schematic of the idealized self-propelled swimmer with a combined heaving and pitching hydrofoil. Dashed line represents a virtual body, not present in the computational domain, but its effect is a drag force, $D$, acting on the hydrofoil. The wake element endpoints are colored red and blue for counter-clockwise and clockwise circulation, respectively. (b) Schematic representation of the wetted surface area, $S_w$, (purple shaded area) and the propulsor planform area, $S_p$, (teal shaded area) for a generic swimmer.}
    \label{fig:schematics}
\end{figure}

A combination of a virtual body and a two-dimensional combined heaving and pitching hydrofoil is used to numerically model an intermittent self-propelled swimmer (Figure~\ref{fig:schematics}a).  The virtual body is not present in the computational domain, but its presence is modeled as a drag force,  
\begin{align} \label{eq:drag}
D = \ea{\frac{1}{2}}\, \rho C_D S_w U^2,
\end{align}

\noindent that follows a typical high Reynolds number drag law \ea{that is used for streamlined bodies \cite{Munson1990, Godoy-Diana2018}}.  Here $\rho$ is the density of the fluid, $C_D$ is the drag coefficient, $S_w$ is the combined wetted surface area of the virtual body and the propulsor (Figure~\ref{fig:schematics}b), and $U$ is the swimming speed. \ea{It has been shown that fish follow this drag law when $Re > 10^4$ \cite[]{Gazzola2014} and that flow speed variations due to intermittent swimming have little impact on the net thrust force of a swimmer \cite[]{VanBuren2018}.  Taken together, these studies support the use of this drag law.}  The propulsor has a teardrop cross-sectional shape \cite{marais2012stabilizing} with a thickness-to-chord ratio of $b/c = 0.1$, where the chord length is fixed to $c = 0.1$ m throughout this study.  \ea{The propulsor planform area per unit span is $S_p = c$}.  Throughout the current study, the area ratio is fixed to $S_{wp} = S_w/S_p = 10$\ea{, since typical fish that swim intermittently have area ratios in the range of $8 \leq S_{wp} \leq 12$ \cite{Videler1981, Videler1991, Webber2001}}.  The drag coefficient and area ratio can be rewritten as the Lighthill number \cite{Eloy2012,Moored2018a}, 
\begin{equation}\label{eq:Lighthill}
Li = C_D S_{wp},
\end{equation}

\noindent which characterizes how the body and propulsor geometry affects the balance of thrust and drag forces on a swimmer.  High $Li$ leads to low self-propelled swimming speeds and \textit{vice versa}. In the drag law, $C_D$ is fixed regardless of whether the swimmer is in a bursting or coasting phase.  Consequently, the $Li$ is also fixed throughout a simulation, which indicates that any intermittent swimming benefits that are observed are solely due to inviscid mechanisms \cite{Akoz2018}. \ea{Previous biological and experimental studies report a range of drag coefficients of swimming fish from 0.001 to 0.05 depending upon the $Re$, body shape, and typical gait of a fish \cite{Videler1981, Webb1984, Barrett1999, Ehrenstein2014}. Therefore, we examined a range of $Li$ that is representative of what is seen in the nature, namely $Li = 0.05$, $0.1$, $0.2$}. 

The virtual body is also given a mass, $m$, that is non-dimensionalized with the characteristic added mass of the propulsor as, $m^* = m/ \rho S_p c$. The dimensionless mass is set to $m^* = 3.86$, which is typical of fish that swim intermittently \cite{Akoz2018}.  
\begin{figure}[t]
\centering
\includegraphics[width=0.90\textwidth]{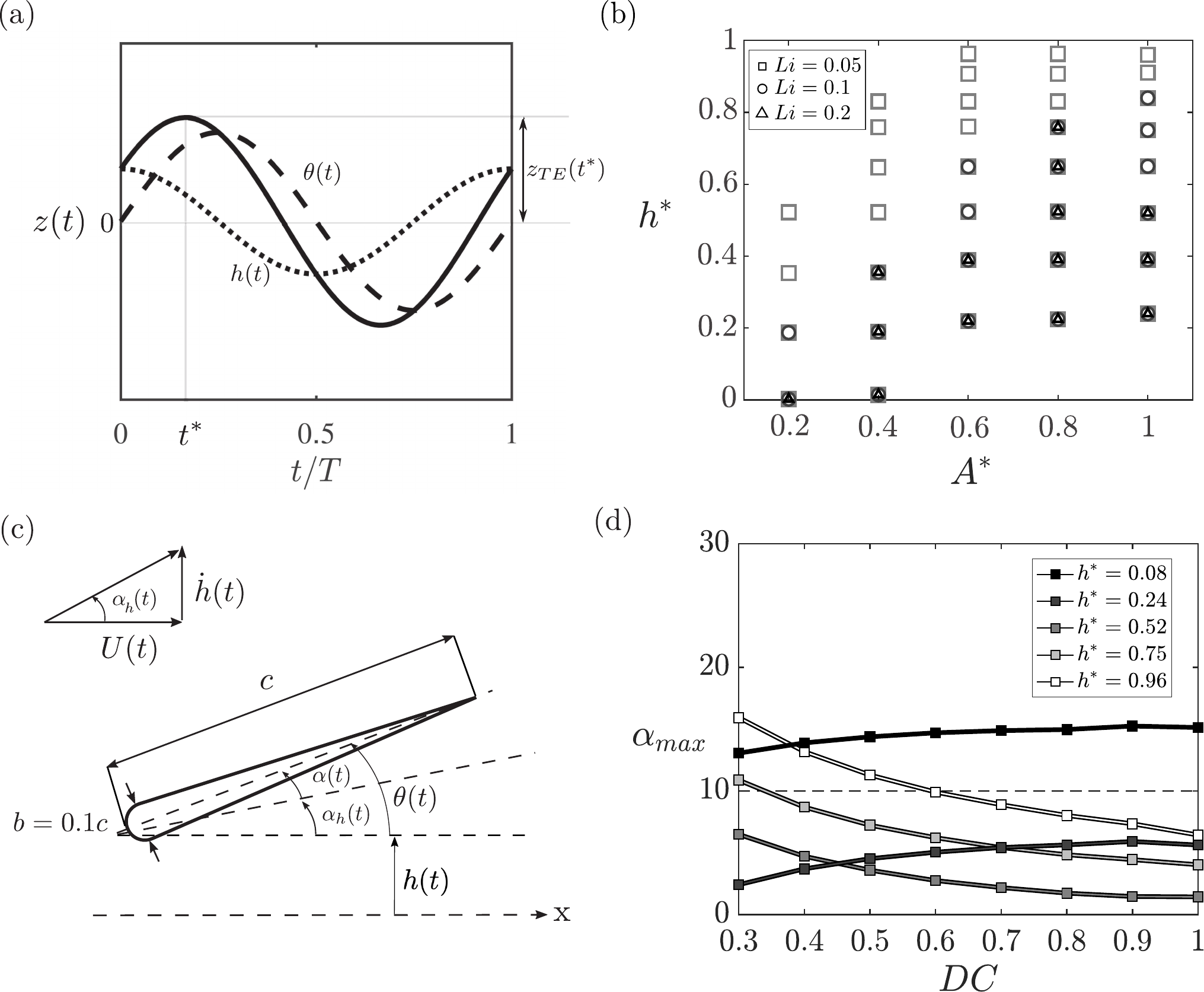}
\caption{(a) The trailing edge amplitude of motion, $z_{TE}(t)$, decomposed into its heaving, $h(t)$, and pitching, $\theta(t)$, components.  (b) Variable range for the study shown as a function of dimensionless amplitude and \ea{heave ratio}. The Lighthill number of a simulation is denoted by different marker types with $Li=0.05$, $0.1$, and $0.2$ represented by square, circle, and triangle markers, respectively. (c) Schematic of the variables affecting the angle of attack. The chord length is, $c=0.1$ m, the maximum thickness of the airfoil is 10\% of the chord length, $U(t)$ is the swimming speed, $\dot{h}(t)$ is the heave velocity, and $\alpha_h$ is the angle of attack from the heave component of the motion. (d) Maximum angle of attack as a function of duty cycle for $Li = 0.05$.  The dashed line represents the cut off maximum angle of attack at \ea{$\alpha_\text{max} = 10^o$}}\label{fig:AoA}
\end{figure}

The hydrofoil undergoes combined heaving and pitching motions, 
\begin{equation}
  h(t) = h_0 \sin(2\pi ft) \qquad \theta(t) = \theta_0 \sin(2\pi ft+\phi) \label{eq:kinematics} 
\end{equation}

\noindent where the pitch axis is located at the leading edge.  Here the frequency of motion is $f$, the heave and pitch amplitudes are $h_0$ and $\theta_0$, respectively, the time is $t$, and the phase difference between the heaving and pitching motions is $\phi$, which is fixed at $\phi = -\pi/2$ where high efficiency is expected \cite{Read2003}.  The time-varying $z$ position of the trailing edge is defined as, 
\begin{equation}
    z_{TE}(t) = h(t) + c  \sin\left[\theta(t)\right].
    \label{eq:atstar}
\end{equation}

\noindent The maximum of this $z$ position occurs at time $t^*$ (Figure \ref{fig:AoA}a).  This time is then used to determine the peak-to-peak trailing edge amplitude,
\begin{align} \label{eq:amplitude}
    A = 2 z_{TE}(t^*). 
\end{align}

\noindent The dimensionless amplitude is defined as $A^*=A/c$. \ea{Once two variables of the heave amplitude, pitch amplitude and the total amplitude are prescribed, the other can be determined. In the present study, we set the maximum pitch amplitude, $\theta_0$, and the total amplitude, $A$, while $h_0$ and $t^*$ are solved for iteratively by using the combination of Equations (\ref{eq:kinematics}) - (\ref{eq:amplitude}).}  The proportion of the peak-to-peak amplitude that comes from the heaving and pitching motions, respectively, are 
\begin{equation}
h^*=\frac{2h(t^*)}{A} \qquad \mbox{and} \qquad \theta^* =\frac{ 2 c  \sin\left[\theta(t^*)\right]}{A} \label{eq:normalizedheavepitch}
\end{equation}

\noindent When $h^* = 1$ the hydrofoil is purely heaving, when $h^* = 0$ the hydrofoil is purely pitching, and when $h^* = 0.5$ the hydrofoil is undergoing a combined heaving and pitching motion with half of the trailing edge amplitude coming from the heaving motion and the other half coming from the pitching motion.  In this sense, $h^*$ is the dimensionless heave ratio and $\theta^*$ is the dimensionless pitch ratio, which are related by the identity,  
\begin{equation}
h^* + \theta^* = 1
\end{equation}

\noindent The selected test domain as a function of $h^*$, $A^*$ and $Li$ are shown in Figure \ref{fig:AoA}b.  The maximum angle of attack of the hydrofoil is defined as (Figure \ref{fig:AoA}c),
\begin{align}
\alpha_{\text{max}} = \text{max}\, \left|\theta(t) - \arctan\left[\frac{\dot{h}(t)}{U\ea{(t)}}\right]\right|
\end{align}

\begin{figure}[t]
\centering 
\includegraphics[width=0.85\textwidth]{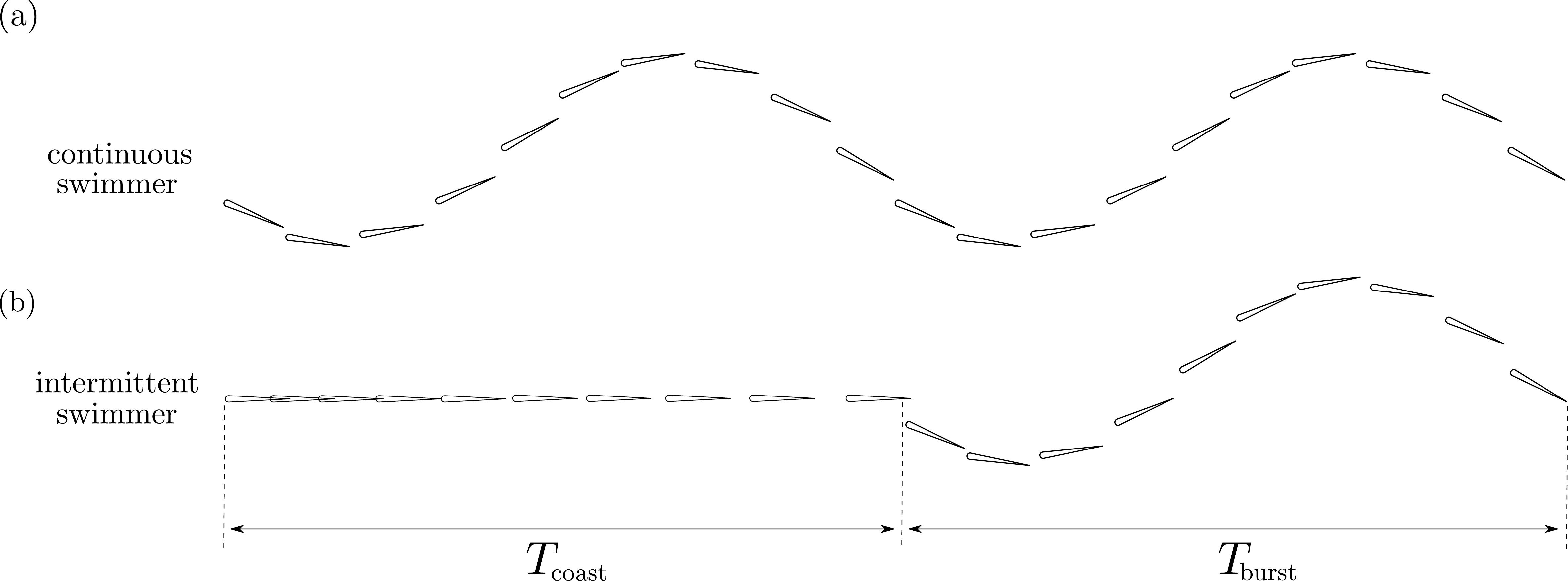}
\caption{ Snapshots of (a) two oscillation cycles for a hydrofoil continuously swimming to the left, and (b) one oscillation cycle of a hydrofoil intermittently swimming to the left with $DC=0.5$. }\label{fig:intermit}
\end{figure}

\noindent When the angle of attack is \am{$\alpha \geq 10^o$} flow separation from the leading edge of a flapping foil becomes significant \cite{Akbari2003}. Since the potential flow method in this study does not model leading-edge separation, we restrict the maximum angle of attack to \am{$\alpha_\text{max} < 10^o$} (Figure \ref{fig:AoA}d; see Section \ref{s:valid} for further discussion).  The intermittency of motion is controlled through the duty cycle,  
\begin{align}
    DC = \frac{T_\text{burst}}{ T_\text{cycle}},
\end{align} 
\noindent where bursting period is $T_\text{burst}$, and the total cycle period is $T_\text{cycle} = T_\text{burst} + T_\text{coast}$, which is simply the addition of the bursting and coasting periods.   The bursting period is inversely related to the frequency of motion as $T_\text{burst} = 1/f$ and the total cycle period is determined by $T_\text{cycle} = T_\text{burst}/DC$.  In the current study the duty cycle ranges from $0.3 \leq DC \leq 1$ in $0.1$ increments.  Snapshots of two oscillation cycles for a continuously swimming hydrofoil and one cycle for an intermittently swimming hydrofoil are shown in Figure \ref{fig:intermit}. \ea{The combined heaving and pitching motion for an intermittent swimmer is defined as follows, 
\begin{equation}
   h(\tau) = \left\{
  \begin{array}{lr}
    y_s(\tau)h_0\sin(2\pi \tau) & : 0\le \tau \le 1\\
    0 & : 1 \le \tau \le \frac{1}{DC},
  \end{array}
\right.\label{eq:heaveDC}
\end{equation}
\begin{equation}
   \theta(\tau) =\left\{
  \begin{array}{lr}
    y_s(\tau)\theta_0\sin(2\pi \tau+\phi) & : 0\le \tau \le 1\\
    0 & : 1 \le \tau \le \frac{1}{DC},
  \end{array}
  \right.\label{eq:pitchDC}
\end{equation}

\noindent where $\tau=t/T_{\text{burst}}$ and
\begin{equation}
    y_s(\tau)=
  \begin{array}{lr}
   -\tanh(\beta \tau) \tanh\left[\beta (\tau-1)\right].
  \end{array}
  \label{eq:rampupfunc}
\end{equation}
\noindent Equations~\eqref{eq:heaveDC} and \eqref{eq:pitchDC} define a reference signal where $0 \leq \tau \leq 1/DC$.} The signal used in the simulations has $N_\text{cyc}$ repetitions of this reference signal.  In order to obtain discretization independent solutions as the time step size is reduced, the discontinuous lateral velocity and consequently the accelerations at the junction of the burst phase and coast phase must be smoothed. \ea{To do this, a dimensionless smoothing function, $y_s(\tau)$, is multiplied with the lateral displacements as defined in equation \eqref{eq:rampupfunc}.} This function modifies the slope of the total lateral displacement at $t/T_\text{burst}=0$ and $t/T_\text{burst}=1$ to ensure a desingularized smooth junction with the coast phase where $\beta$ controls the radius of curvature of the junction. In the current study, \ea{$\beta = 30$ is chosen since it does not significantly alter the burst signal from a sine wave and it does not require a large number of time steps per burst period to reach a converged solution}. Additionally, if $DC = 1$, then the signal reverts to a continuous sinusoidal signal.

Lastly, decreasing the duty cycle decreases the average velocity, which can either increase or decrease the maximum angle of attack depending upon $h^*$ as shown in Figure \ref{fig:AoA}d. Therefore, some low $DC$ results were eliminated from the data set when the maximum angle of attack was above \am{$10^o$}. 

\subsection{Numerical Approach}
An unsteady two-dimensional boundary element method based on potential flow theory is employed to calculate the flow field around self-propelled hydrofoils. The flow is assumed to be irrotational (except on the boundary elements), incompressible and inviscid, such that the velocity can be defined as $\textbf{u}=\nabla{\phi^*}$ where $\phi^*$ is the perturbation potential in the inertial frame. The incompressible continuity equation is then reduced to Laplace's equation, $\nabla^2 \phi^* = 0$, which governs the fluid flow. There is a general solution to Laplace's equation in the form of a boundary integral equation that is used to determine the potential field and the flow field produced by a body and its wake.

To solve this problem numerically, constant strength source and doublet line elements are distributed over the body and the wake. Each body element is assigned a collocation point, which is located at the center of each element and shifted 1\% of the local thickness into the body where a constant-potential condition is applied to enforce no flux through the surface (i.e. Dirichlet formulation). This results in a matrix representation of the boundary condition that can be solved for the body doublet strengths once a wake shedding model is applied. Additionally, at each time step, a wake boundary element is shed with a strength that is set by applying an explicit Kutta condition, where the vorticity at the trailing edge is set to zero \citep{Willis2006, Wie2009,Pan2012}.  A wake roll-up algorithm is implemented at each time step where the wake elements are advected with the local velocity.  During wake roll-up, the point vortices, representing the ends of the wake doublet elements, must be desingularized for numerical stability of the solution \citep{Krasny1986}.  At a cutoff radius of $\epsilon/c=5\times10^{-2}$, the irrotational induced velocities from the point vortices are replaced with a rotational Rankine core model. The tangential perturbation velocity component is calculated by local differentiation of the perturbation potential. Finally, the pressure acting on the body is found via applying the unsteady Bernoulli equation and the forces acting on the body are determined by integrating the pressure over the hydrofoil boundary. 

To examine intermittent motion, the free swimming condition is enforced through a single degree of freedom equation of motion that allows the streamwise translation of the hydrofoil. Following \citet{Borazjani2009}, the velocity of the swimmer at the $(n+1)^{th}$ time step is calculated through forward differencing and the position is calculated by using the trapezoidal rule,

\begin{align}
U^{n+1} = U^n + \frac{F^n_{x,net}}{m}\Delta t \\
x^{n+1}_{LE} = x^n_{LE} + \frac{1}{2}(U^{n+1} + U^n)\Delta t
\end{align}

\noindent where $F^n_{x,net}$ is the net force acting on the hydrofoil in the streamwise direction at the $n^{th}$ timestep, $x_{LE}$ is the leading edge position of the hydrofoil. The two-dimensional formulation in the current study has been validated extensively against continuous swimming theory, numerics and experiments \cite{Quinn2014,Moored2018Bem,kurt2019swimming}. More details about the numerical scheme can be found in these references \cite{Katz1991, Moored2018Bem}. 

\subsection{Output Variables}
The cycle averaged variables of interest include the mean velocity, $\overline{U}$, thrust, $\overline{T}$, and power, $\overline{P}$, which are calculated once the swimmers have reached a cycle-averaged steady-state swimming condition. The computations are considered to be at this steady state when there is negligible mean net thrust acting on the swimmer, defined as,
\begin{align}
C_{T,\text{net}} = \frac{\overline{T} - \overline{D}}{\rho S_p \overline{U}^2} < 10^{-4}.
\end{align}

\noindent The thrust is calculated as the streamwise-directed pressure forces while the power input to the fluid is calculated as the negative inner product of the element force vector and its velocity vector, that is, $P = -\int_\mathcal{S} \mathbf{F}_{\text{ele}} \cdot \mathbf{u}_{\text{ele}} \, d\mathcal{S}$ where $\mathcal{S}$ is the body surface. The thrust and power coefficients are defined as, 
\begin{align}\label{eq:coeffs}
C_T = \frac{\overline{T}}{\rho S_p f^2 A^2}, \qquad  C_P = \frac{\overline{P}}{\rho S_p f^2 A^2 \overline{U}}
\end{align}

\noindent The propulsive efficiency, $\eta$, and the ratio of the propulsive efficiencies of the intermittent swimmer and the continuous swimmer, $\eta^*$, are defined as,
\begin{align}
\eta = \frac{C_T}{C_P}, \qquad  \eta^* = \frac{\eta_\text{int}}{\eta_\text{cont}}.
\end{align}

\noindent Now, when $\eta^*>1$ intermittent swimming is more efficient than the continuous swimming, when $\eta^*=1$ continuous and intermittent swimming have the same efficiency, and when $\eta^*<1$, continuous swimming is more efficient.

In this study the simulations are self-propelled and it is important to consider how the thrust and power coefficients connect to the swimming speed and range of the swimmer.  Once the steady-state of the cycle-averaged swimming speed is reach then the time-averaged thrust and drag balance, $\overline{T} = \overline{D}$.  By substituting the drag law (\ref{eq:drag}), the thrust coefficient definition (\ref{eq:coeffs}), and the Lighthill number definition (\ref{eq:Lighthill}), the swimming speed is,
\begin{align}\label{eq:speed}
\overline{U} \approx fc \, \mathcal{U}^*, \qquad \mbox{where} \qquad \mathcal{U}^* = \sqrt{\frac{2 {A^*}^2 C_T}{Li}}.
\end{align}

\noindent \ea{Here, the approximation of $\overline{U^2} \approx \overline{U}^2$ has been used for simplicity.  For the selected range of data in the current study, this approximation is within 1.5\% of the exact value.}  The variable $\mathcal{U}^*$ represents the dimensionless swimming speed and if the frequency and chord length of the swimmer are fixed, as they are throughout this study, then increases or decreases in the dimensionless speed directly translate into increases or decreases in the dimensional speed.  Now it becomes clear that increases in the thrust coefficient do not necessarily translate to higher swimming speeds if the dimensionless amplitude or Lighthill number are changed as in this study.  

The range of the swimmer is simply calculated as the amount of energy available multiplied by the distance the swimmer travels per unit energy, $\mathcal{R} = E_\text{avail}\left(\overline{U}/\overline{P}\right)$.  By substituting in the power coefficient definition (\ref{eq:coeffs}) the range equation is expanded,
\begin{align}\label{eq:range}
\mathcal{R} = \frac{E_\text{avail}}{\rho S_p f^2 c^2} \, \mathcal{R}^*, \qquad \mbox{where} \qquad \mathcal{R}^* = \frac{1}{{A^*}^2 C_P}.
\end{align}

\noindent Increases or decreases in the dimensionless range, $\mathcal{R}^*$, directly increase or decrease the dimensional range as long as the frequency, chord length, propulsor area, fluid density and energy available remain constant as in the current study.  Similar to the thrust coefficient, interpreting changes in the power coefficient as changes in the range of the swimmer is confounded when the dimensionless amplitude is varying.  Therefore, in this study we will report the dimensionless speed and range instead of the typical thrust and power coefficients.

%-------------------------------------------------RESULTS-----------------------------------------------------%
\subsection{Discretization Independence and Validation} \label{s:valid}
\begin{figure}[h!]
    \centering
    \includegraphics[width=0.85\linewidth, scale=0.75]{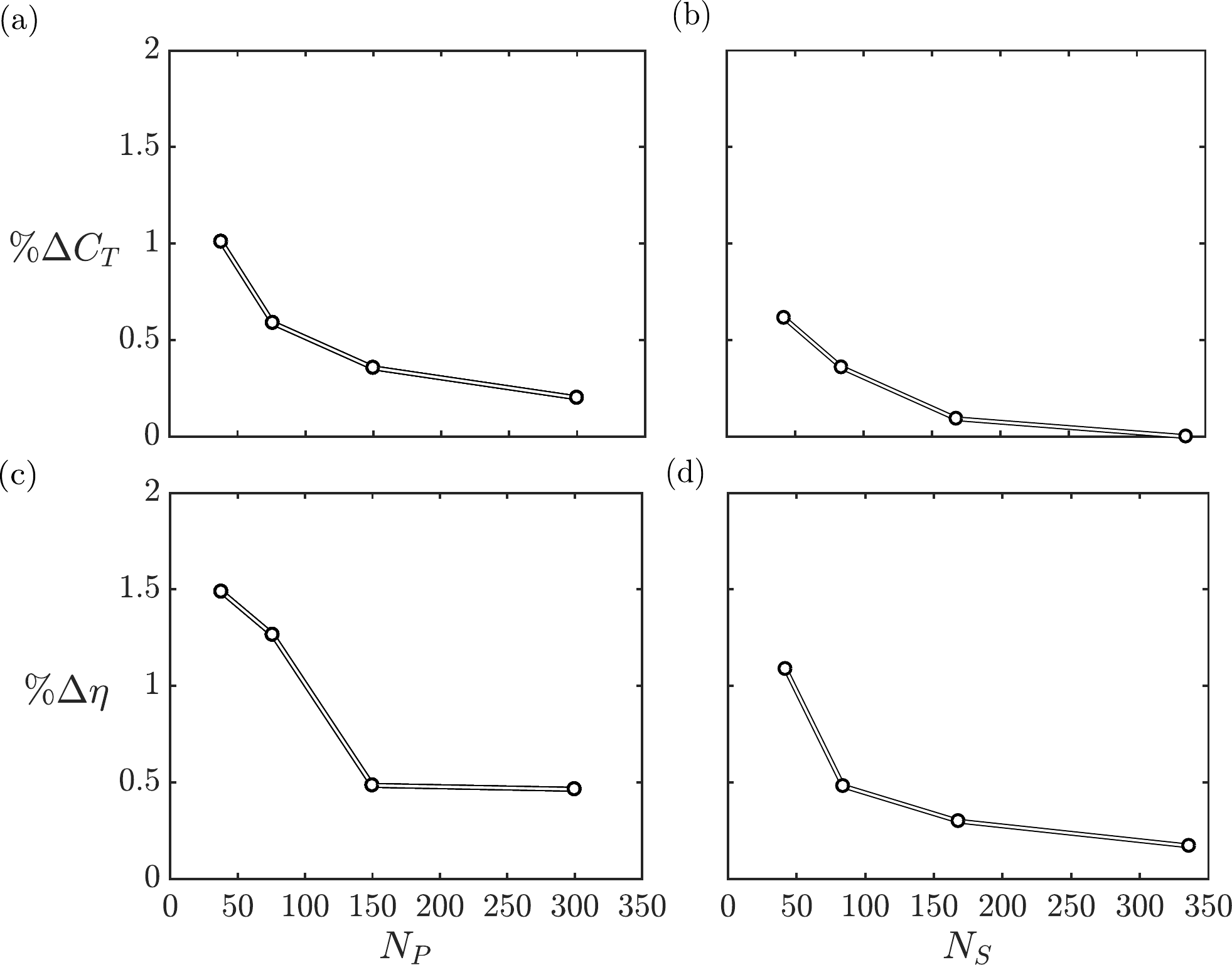}
    \caption{\ea{Discretization independence graphs showing the percent change in (a) the thrust coefficient and (c) efficiency when the number of time steps are held constant ($N_S=168$) and the number of body elements are doubled. Similarly, discretization independence graphs showing the percent change in (b) the thrust coefficient and (d) efficiency when the number of body elements are held constant ($N_P=150$) and the number of time steps per bursting period are doubled.}}
    \label{fig:converge}
\end{figure}

A discretization independence study was performed to determine the necessary resolution for the simulations.  A self-propelled combined heaving and pitching hydrofoil swimming intermittently was simulated with $h^*=0.27$, $Li=0.05$, $m^*=3.863$, $f=1.0$ Hz, $DC=0.5$, and $A^* = 0.4$.  \ea{More extensive discretization independence simulations at $A^* = 1$ are presented in Appendix \ref{App conv}}.  For the $A^* = 0.4$ case, the hydrofoil started with a small initial velocity and was simulated for fifty cycles of swimming with the time-averaged data obtained by averaging over the last cycle.  \ea{The steady-state criterion was achieved within fifty cycles.}  Figure \ref{fig:converge} presents the percent change in the thrust coefficient and the efficiency when either the number of body elements ($N_P$) or the number of time steps \ea{over a burst period} ($N_S$) are doubled.  It was determined that the mean thrust and swimming efficiency changed by less than 1\% when the number of body elements or the number of times steps per cycle were doubled for $N_P \geq 150$ and $N_S \geq 168$, respectively.  \ea{When considering the more extensive independence study (Appendix \ref{App conv}), the thrust and efficiency change by less than approximately 2\% when the number of body elements and time steps per bursting period are doubled from $N_P=150$ and $N_S=168$, respectively.}  Therefore, for all subsequent data presented in this study $N_P=150$ and $N_S=168$. 

\begin{figure}[h!]
    \centering
    \includegraphics[width=0.98\linewidth, scale=0.75]{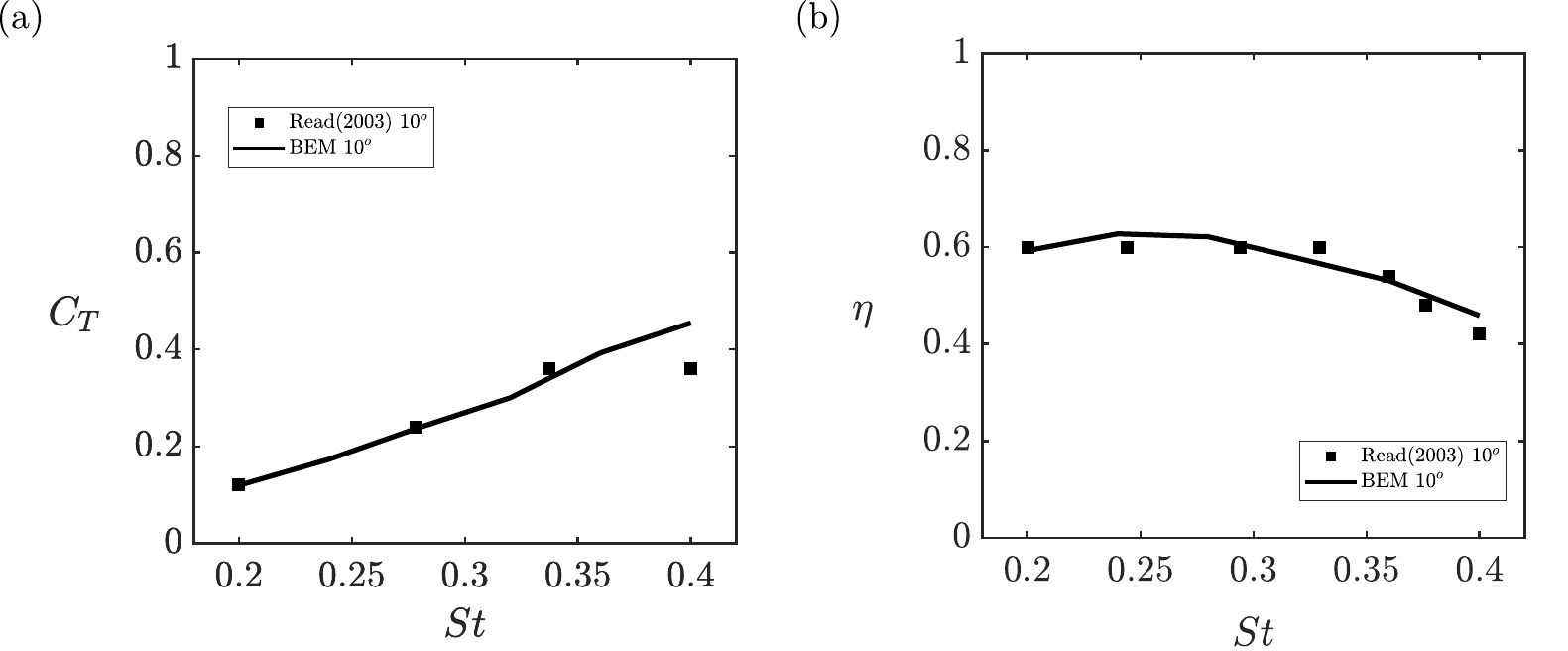}
    \caption{\am{Comparison of the BEM potential flow simulations to the experiments of \cite{Read2003}. The comparison is shown for the angle of attack of $\alpha_\text{max} =10^o$ with $h_0/c = 0.75$ and $\psi = 90^o$, which is the phase difference between pitch and heave. Presented is (a) the thrust coefficient and (b) efficiency as functions of the Strouhal number.  The lines and markers represent the BEM numerical solutions and the experimental data, respectively.}}
    \label{fig:valid}
\end{figure}

Previously, the numerical methodology and solver were validated using numerous canonical steady and unsteady theories, computations and experiments for pitching, heaving, and undulating foils \cite{Quinn2014,Moored2018Bem,kurt2019swimming}.  However, combined heaving and pitching motions were not validated, which is especially important since the current form of the numerical solution does not account for leading-edge separation; a well-known phenomenon for high angle-of-attack combined heaving and pitching hydrofoils \cite{Anderson1998}.  The simulations in the current study were restricted to have maximum angles of attack of $\alpha_\text{max} \le 10^o$ since for angles of attack greater than $10^o-15^o$, leading-edge vortex formation has a significant effect on the thrust and power in experiments \cite{Pan2012}, and the current form of the numerical solution does not account for it.  \am{Figure \ref{fig:valid} presents the thrust coefficient and efficiency as functions of the Strouhal number for both the current numerical solution and the experiments from \cite{Read2003}.  The comparison is for a case with the maximum angle of attack of $\alpha_\text{max} = 10^o$. These validation simulations were not self-propelled, but performed at a fixed velocity and a drag force with a drag coefficient of $C_D = 0.05$ was added to the potential flow solution since this was the approximate profile drag assumed in the experiments \cite{Read2003}.  All of the parameters and variables were matched to the experimental work with the $h_0/c = 0.75$ case used here.  Figure \ref{fig:valid}a shows that thrust from the simulations match the experiments quite well for Strouhal numbers from $0.2 \leq St \leq 0.4$. Figure \ref{fig:valid}b shows that for $\alpha_\text{max} = 10^o$ the efficiency calculated in the simulations match quite closely to the experiments with a maximum error of 6.7\%. However, for greater angles of attack the error would climb further, necessitating the use of a leading-edge separation model to produce accurate results \cite{Pan2012}.  Instead, the maximum angle of attack was limited $\alpha_\text{max} \leq 10^o$ where the potential flow simulations are in good agreement with high Reynolds number ($Re = 4 \times 10^4$) experiments.}

\section{Results} \label{s:results}
\subsection{Continuous Swimming with Combined Heaving and Pitching Motions}
\begin{figure}[h]
\centering 
\includegraphics[width=0.75\textwidth,scale=0.75]{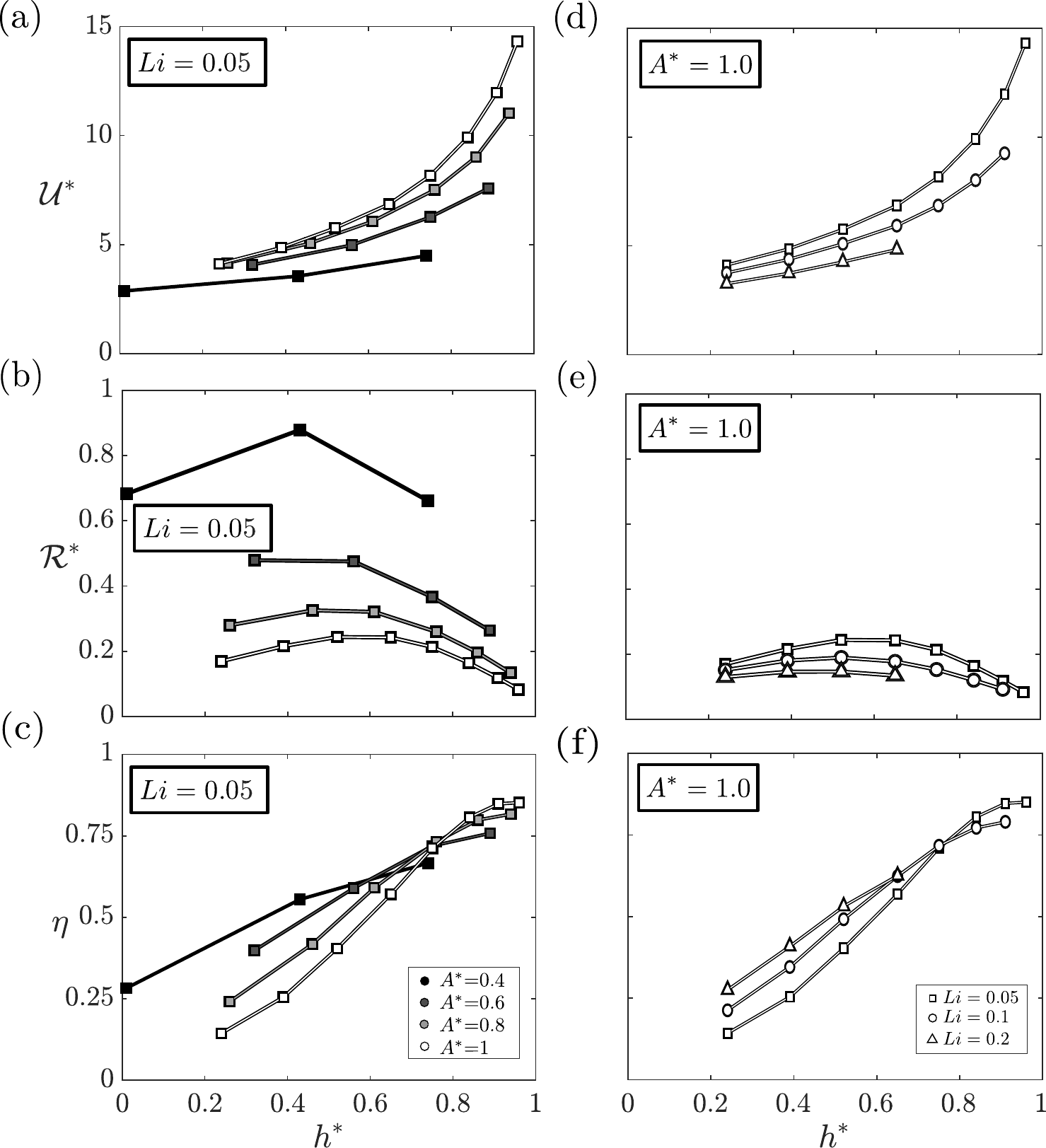}
    \caption{(a) - (c) Dimensionless speed, dimensionless range and efficiency as a function of $h^*$ at $Li=0.1$ with varying $A^*$. (d) - (f) Dimensionless speed, dimensionless range and efficiency as a function of $h^*$ at $A^* = 1.0$ and varying $Li$.}
    \label{fig:contnuous_swimming_performance}
\end{figure}

Figure \ref{fig:contnuous_swimming_performance}a - \ref{fig:contnuous_swimming_performance}c presents the dimensionless speed, dimensionless range, and the efficiency as a function of $h^*$ for various dimensionless amplitudes at \am{$Li = 0.05$}.  The amplitudes ranging from low to high are denoted with marker and line colors ranging from black to white, respectively.  The dimensionless speed increases monotonically as the dimensionless \ea{heave ratio} is increased for a fixed $A^*$.  As $A^*$ increases for a fixed $h^*$, the dimensionless speed also increases.  In contrast, the dimensionless range increases for decreasing $A^*$.  \ea{For a fixed $A^*$ the dimensionless range is maximized around $0.4 \leq h^* \leq 0.6$ with the optimal $h^*$ increasing with decreasing $Li$}.  The efficiency shows a more complicated trend where for $0 \leq h^* \leq 0.5$ lower $A^*$ motions are found to be more efficient, while for  $0.5 < h^* \leq 1$ higher $A^*$ motions are more efficient.  There is also an optimal $h^*$ that maximizes the efficiency in the range $0.5 < h^* \leq 1$.
\begin{figure}
    \centering
    \includegraphics[width=0.90\textwidth]{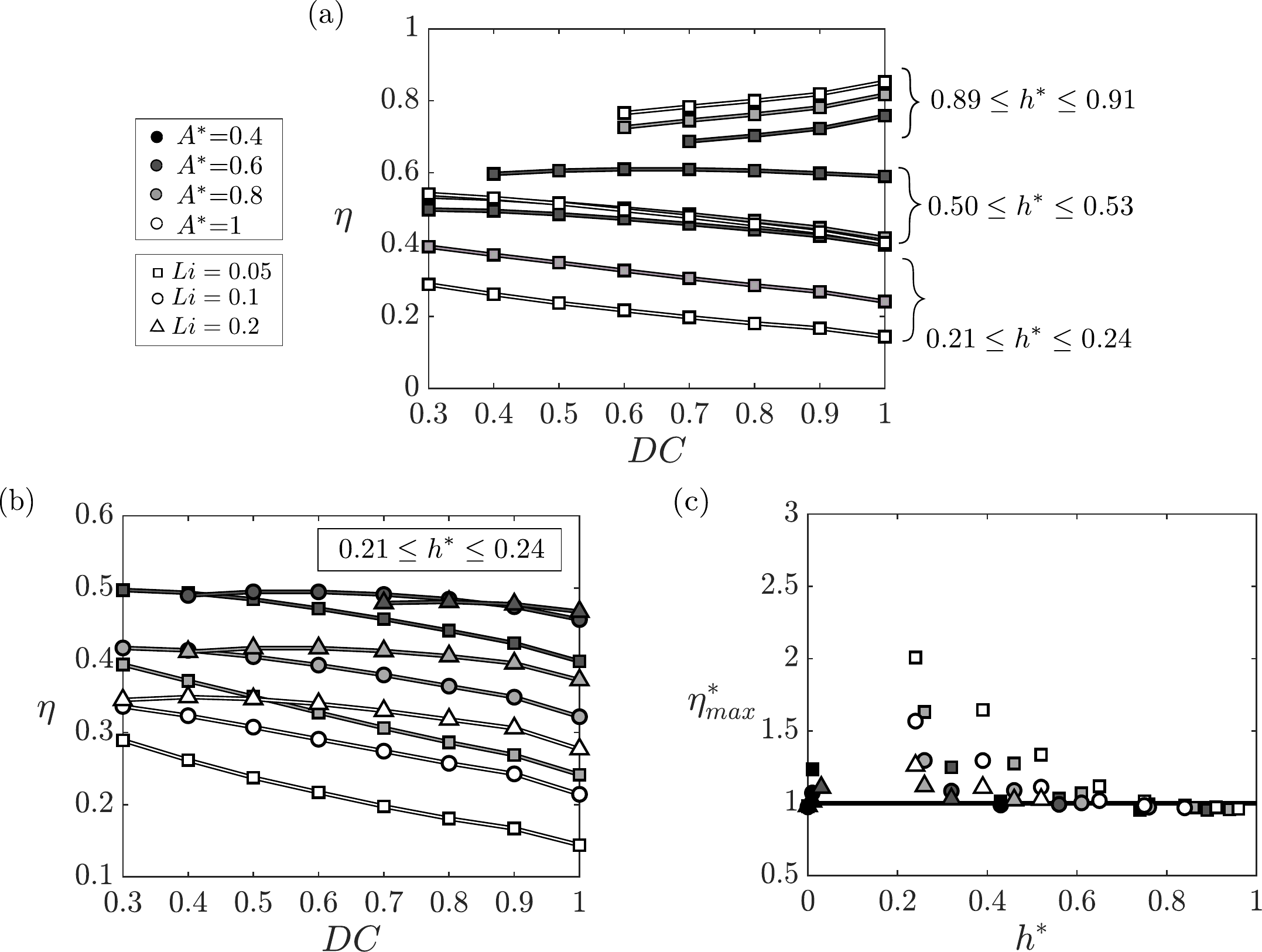}
    \caption{(a) Efficiency of intermittent swimmers for varying duty cycles, amplitudes and \ea{heave ratios} with $Li = 0.05$. The efficiency curves are grouped for high, medium and low $h^*$.  (b) Efficiency of intermittent swimmers for varying duty cycles, amplitudes and Lighthill numbers with $0.21 \leq h^* \leq 0.24$.  (c) The ratio of the maximum efficiency of an intermittent swimmer to its continuous swimming efficiency as a function of the \ea{heave ratio}. }
    \label{fig:intermittent_performance}
\end{figure}

Figure \ref{fig:contnuous_swimming_performance}d - \ref{fig:contnuous_swimming_performance}f presents the dimensionless speed, dimensionless range, and the efficiency as a function of $h^*$ for various $Li$ at a fixed amplitude of $A^* = 1.0$.  Different $Li$ are denoted by the different marker styles.  When the $Li$ is decreased this leads to higher dimensionless speeds, dimensionless ranges, and peak efficiencies.  \ea{However, the efficiency in the range $0 \leq h^* \leq 0.5$ increases with increasing $Li$}.  Lowering $Li$ occurs when the drag on the body and propulsor of the swimmer is reduced and in turn the thrust that is produced to balance the drag can also be reduced, leading to more efficient motions.  Interestingly, the optimal range and optimal efficiency do not occur at the same $h^*$ (as in the case of varying $A^*$), and as $Li$ decreases both optimal points move to higher $h^*$ values.

\subsection{Intermittent Swimming with Combined Heaving and Pitching Motions}
\ea{Figure \ref{fig:intermittent_performance} extends the efficiency results shown in Figure \ref{fig:contnuous_swimming_performance} to intermittent swimming. To systematically investigate the effect of $DC$ on the performance, the $Li$ was fixed in Figure \ref{fig:intermittent_performance}a and $h^*$ is nearly fixed, varying over a narrow range in \ref{fig:intermittent_performance}b. Finally, in Figure \ref{fig:intermittent_performance}c, the most efficient intermittent swimmer (optimal $DC$ for a given $h^*$, $A^*$ and $Li$) was compared against its continuous swimming counterpart.} 

Figure \ref{fig:intermittent_performance}a presents the efficiency as a function of the duty cycle for various dimensionless amplitudes and at a fixed Lighthill number of $Li=0.05$. These data are grouped into pitch-dominated motions \ea{($0.21\leq h^* \leq 0.24$)}, heave-dominated motions ($0.89\leq h^* \leq 0.91$), and motions with nearly equal pitching and heaving amplitudes ($0.5\leq h^* \leq 0.53$).  As in the case of continuous swimming, for pitch-dominated motions the efficiency is increased for decreasing dimensionless amplitudes, while for heave-dominated motions the efficiency is increased for increasing dimensionless amplitudes.  For nearly equal pitching and heaving amplitude cases, \ea{there is no clear trend between the efficiency and $A^*$ }.  As the duty cycle is decreased from $DC = 1$ (continuous swimming) to $DC = 0.3$ (highly intermittent swimming), the efficiency increases for pitch-dominated motions while it decreases for heave-dominated motions.  For $0.5\leq h^* \leq 0.53$ \ea{the efficiency increases as the duty cycle decreases, reaches a maximum and starts to decrease for lower duty cycles.} This data indicates that intermittent swimming can only improve efficiency when a swimmer does not employ heave-dominated swimming motions.

Figure \ref{fig:intermittent_performance}b presents the efficiency as a function of duty cycle for a narrow range of the heave ratio of $0.21 \leq h^* \leq 0.24$ for various dimensionless amplitudes and Lighthill numbers. In this pitch-dominated regime with $h^* < 0.5$, it is observed that \ea{peak efficiencies occur for intermittent swimming ($DC < 1$) and when $A^*$ is fixed and $Li$ decreases the optimal $DC$ decreases. In this regime, the largest percent increases in efficiency occur for the highest amplitude and lowest $Li$ cases, as observed previously for pure pitching motions \cite{Akoz2018}}.

Figure~\ref{fig:intermittent_performance}c presents the normalized maximum efficiency of intermittent swimming, $\eta^*_\text{max}$, as a function of the dimensionless \ea{heave ratio}.  For a given ($h^*,\, Li,\, A^*$) the maximum efficiency of intermittent swimming is the highest efficiency that occurs when $DC < 1$ (intermittent swimming).  This efficiency is then normalized by the efficiency when $DC = 1$ (continuous swimming), such that $\eta^*_\text{max} = \eta_\text{max}/\eta_\text{cont}$. If this normalized efficiency is greater than, equal to, or less than one then intermittent swimming is more efficient, equally efficient, or less efficient, respectively.  For combined heaving and pitching motions, intermittent swimming is energetically advantageous for $h^* < 0.7$ with greater than \am{105\%} increase in the efficiency observed for the lowest $h^*$ values examined.  When $h^*\geq 0.7$, continuous swimming is advantageous and peak efficiencies are observed with values as high as 85\%.  Furthermore, at a fixed \ea{heave ratio} below $h^* = 0.7$, the maximum normalized efficiency increases with increasing $A^*$ and decreasing $Li$. Therefore, the greatest energetic benefits of intermittent swimming over continuous swimming can be derived when a swimmer: (i) uses low to moderate \ea{heave ratio}s ($h^* < 0.7$), (ii) has a low Lighthill number, and (iii) swims with a high amplitude of motion.

\subsection{Inviscid Mechanisms behind the Efficiency of Continuous and Intermittent Swimming}
To gain deeper insight into the inviscid mechanisms underlying these basic ``rules'' stated above we can examine Garrick's classic linear theory \cite{Garrick1936} that describes the thrust production and power consumption of a combined heaving and pitching foil.  \ea{Garrick's theory assumes the standard assumptions of potential flow theory where the flow is inviscid, incompressible, and irrotational, and it further assumes that the wake is non-deforming and planar, and that the motion is continuous, harmonic, and of small amplitude.  These assumptions may seem to preclude this theory from capturing the basic physical mechanisms of heaving and pitching \textit{intermittent} swimming, however, previous work \cite[]{Akoz2018} establishes the fact that the thrust and power coefficients from an intermittent swimmer can be transformed into equivalent \textit{continuous} swimming data by dividing the thrust and power coefficients by the duty cycle. In this way the thrust and power coefficients in Garrick's theory can be generally thought of as the thrust and power coefficients averaged only over the bursting portion of the cycle as $C_{T,\text{b}} = C_T/DC$ and $C_{P,\text{b}} = C_P/DC$.  Since the efficiency is the ratio of the thrust and power coefficients then the efficiency does not need to be altered by the duty cycle to apply to intermittent swimming.  This transformation has also be corroborated in experiments \cite[]{floryan2017} by showing the collapse of intermittent thrust and power data to a single curve when scaled by the duty cycle as in the above equations.}

When the pitch axis is about the leading edge and pitching lags the heaving motion by $\phi = \pi/2$ then the thrust and power coefficients from Garrick's theory are,
\begin{eqnarray}
    \label{eq:Garrick_thrust_star}
    C_{T,b} = \blue{\underbrace{\frac{3\pi^3}{32} {\theta^*_0}^2}_{C_{T,p}^\text{add}}} 
    + \red{\underbrace{\frac{\pi^3}{8}{\theta^*_0}^2 \left[(F^2 + G^2) \left(\frac{1}{\pi^2 k^2} +\frac{9}{4}\right) -\frac{3F}{2} -\frac{F}{\pi^2 k^2} + \frac{G}{2\pi k} \right]}_{C_{T,p}^\text{circ}}} 
    + \nonumber \\ \red{\underbrace{\frac{\pi^3}{2}\theta^*_0 h^*_0 \left(\frac{G}{2} + \frac{F}{2\pi k} -\frac{F^2 + G^2}{\pi k}\right)}_{C_{T,hp}^\text{circ}}} 
    + \blue{\underbrace{\frac{\pi^3}{2}{h^*_0}^2 \left(F^2 + G^2\right)}_{C_{T,h}^\text{circ}}},
\end{eqnarray}
\begin{eqnarray}
    \label{eq:Garrick_power_star}
     C_{P,b} =  \underbrace{\frac{3\pi^3}{32}{\theta^*_0}^2}_{C_{P,p}^\text{add}}
     + \underbrace{\frac{\pi^3}{16} {\theta^*_0}^2 \left(\frac{3F}{2} + \frac{G}{\pi k} \right)}_{C_{P,p}^\text{circ}}
      + \underbrace{\frac{\pi^3}{4} \theta^*_0 h^*_0 \left(G - \frac{F}{\pi k}\right)}_{C_{P,hp}^\text{circ}} 
      + \underbrace{\frac{\pi^3}{2} {h^*_0}^2 F }_{C_{P,h}^\text{circ}}.
\end{eqnarray}

\begin{figure}
    \centering
\includegraphics[width=0.99\textwidth]{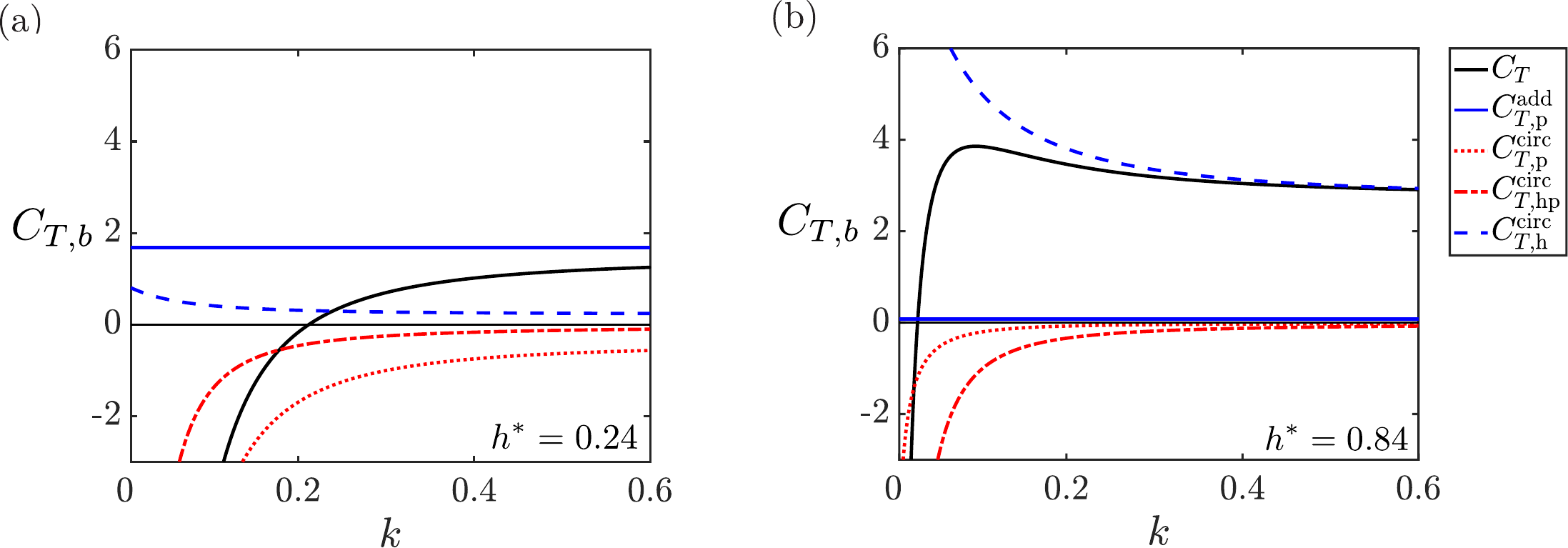}    
\caption{The thrust coefficient as a function of the reduced frequency for (a) $h^* = 0.24$ and (b) $h^* = 0.84$.  The total thrust (black solid line) is decomposed into the pitching added mass term (blue solid line), pitching circulatory term (red dotted line), heaving-pitching circulatory cross term (red dash-dot line), and heaving circulatory term (blue dashed line).}
    \label{fig:CT_decomp}
\end{figure}

\noindent Here, $F$ and $G$ are the real and imaginary parts of Theodorsen's lift deficiency function, respectively \cite{theodorsen1935}, $\theta^*_0 = 2 c \theta_0/A$ and $h^*_0 = 2 h_0/A$, where $\theta_0^* \approx \theta^*$ and $h_0^* \approx h^*$ in the small amplitude limit, \ea{$k$ is the reduced frequency defined as $k= fc/U$}, and the propulsive efficiency is simply $\eta = C_T/C_P$.  The thrust and power are decomposed into their added mass pitching terms, circulatory pitching terms, circulatory heaving-pitching cross terms, and circulatory heaving terms.  The blue and red terms in the thrust coefficient equation represent thrust and drag producing terms, respectively, and match the color scheme in Figure \ref{fig:CT_decomp}.  
\begin{figure}
    \centering
\includegraphics[width=0.99\textwidth]{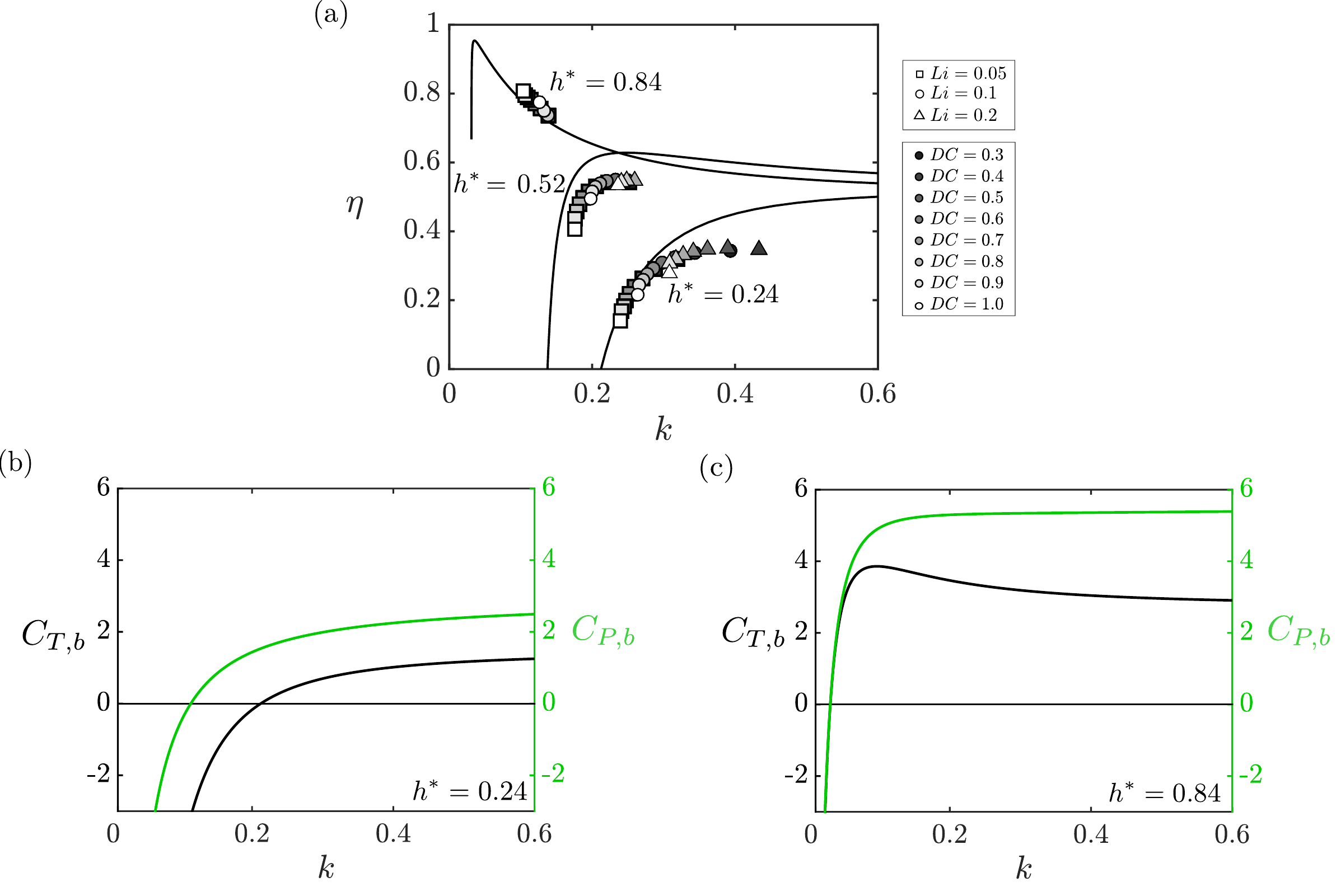}    
\caption{(a) Efficiency as a function of reduced frequency for a variety of $Li$ and $DC$, \am{and for $h^* =0.24$, $0.52$, and $0.84$ with a fixed maximum amplitude of motion of $A^*=1$.  The solid black lines represent Garrick's analytic solutions for the respective $h^*$ values.} Thrust (left axis) and power (right axis) coefficients as functions of reduced frequency for (b) $h^* = 0.24$ and (c) $h^* = 0.84$.}
    \label{fig:efficiencyvsk}
\end{figure}

Figure \ref{fig:efficiencyvsk}a presents the propulsive efficiency as a function of the reduced frequency. \ea{Markers represent the numerical results and solid black lines are the theoretical curves obtained by Garrick's theory for the corresponding $h^*$.}  The data are grouped as pitch-dominated motions with $h^* = 0.24$, as heave-dominated motions with $h^* = 0.84$, and as equally-partitioned heaving and pitching motions with $h^* = 0.52$.  The colors represent the duty cycle with $DC = 1$ to $DC = 0.3$ mapped from white to black, while the marker type indicates the Lighthill number, and the dimensionless amplitude is fixed to $A^* = 1$.  The simulation data are observed to fall along three curves depending upon their dimensionless \ea{heave ratio}.  As the $Li$ increases for a fixed $DC$, the reduced frequency is observed to increase.  Similarly, as the $DC$ decreases for a fixed $Li$, the reduced frequency is also observed to increase.  For the pitch-dominated motions, as the reduced frequency increases there is an increase in the efficiency quite similar to the theoretical trend, though over predicted by the theory.  \ea{For equally-partitioned heaving and pitching motions, the theoretical solution shows that as the reduced frequency increases the efficiency first increases at low $k$, then it reaches a peak value, and finally it decreases at high $k$.  This trend is not observed in the simulation data since it does not extend to a high enough reduced frequency to significantly pass the peak in the theoretical data.}  Still, the theory follows the trends in the simulation data albeit with an over prediction in the efficiency.  For the heave-dominated motions, as the reduced frequency increases the efficiency decreases following the theory quite closely.  This shows that the trends in the data, and therefore the basic physical mechanisms, are reproduced by the theory, and that the theory is more accurate for high $h^*$.  

It is striking that there is a major shift in the trend of the efficiency when going from pitch-dominated motions to heave-dominated motions.  Garrick's theory can explain this shift in terms of the physical nature of the forces that drive the efficiency.  Figure \ref{fig:efficiencyvsk}b and \ref{fig:efficiencyvsk}c present the theoretical solution when $h^* = 0.24$ and $h^* = 0.84$, respectively, for the power and thrust coefficients as a function of reduced frequency.  Both graphs of the power coefficient show a monotonic trend of increasing power with increasing reduced frequency, with the heave-dominated motions using more power than the pitch-dominated motions.  Since the power coefficients show the same basic trend, the qualitative shift in the efficiency trend from pitch-dominated to heave-dominated motions must be due to the thrust production.  Indeed, the thrust coefficient curves are qualitatively different with the pitch-dominated motions showing a monotonically increasing trend with increasing reduced frequency, while for the heave-dominated motions there is a non-monotonic trend with increasing reduced frequency.  This shift in the thrust trend drives the same trends in the efficiency.  

In Figure \ref{fig:CT_decomp} the thrust coefficient is decomposed into its constituent parts from theory into a pitching added mass term (blue solid line), a pitching circulatory term (red dotted line), a heaving-pitching circulatory cross term (red dash-dotted line), and a heaving circulatory term (blue dashed line).  For all motion types where the pitch axis is at the leading edge, the pitching added mass term and the heaving circulatory term always produce thrust, while the pitching circulatory term and the heaving-pitching circulatory cross term always produce drag, even in this potential flow theory.  The drag producing terms cause the total force production to be drag for sufficiently low reduced frequencies, though in the limit as $h^* \xrightarrow{} 1$, the drag terms approach zero.  The major difference between the two motion types is that pitch-dominated motions generate added-mass-based thrust forces with negligible circulatory-based thrust forces while heave-dominated motions generate circulatory-based thrust forces with negligible added-mass-based thrust forces.  When the reduced frequency is increased the wake vortices are stronger and closer to the hydrofoil giving rise to higher downwash/upwash wake effects.   This leads to diminished bound circulation and consequently diminished circulatory-based thrust production.  The added-mass-based thrust forces are constant for all reduced frequencies, since the coefficient has been normalized by a characteristic added mass force.  For heave-dominated motions, the decaying trend in thrust with increasing reduced frequency is driven by the circulatory-based thrust forces, while at low reduced frequencies the drag terms dominate leading to the reversal of the thrust curve and its crossing into net drag.  For pitch-dominated motions, the thrust is asymptotic to the added-mass-based thrust at high reduced frequencies and at low frequencies where the drag terms dominate, the thrust diminishes and crosses into net drag.  

In summary, pitch-dominated motions are driven by added-mass-based thrust production where self-propelled efficiency is maximized by small amplitudes, high Lighthill numbers and low duty cycles, all of which lead to high reduced frequencies.  Heave-dominated motions are driven by circulatory-based thrust production where self-propelled efficiency is maximized by large amplitudes, low Lighthill numbers and high duty cycles, all of which lead to low reduced frequencies.

\subsection{Wake Dynamics}
\begin{figure}[t]
\centering
\includegraphics[width=0.95\textwidth]{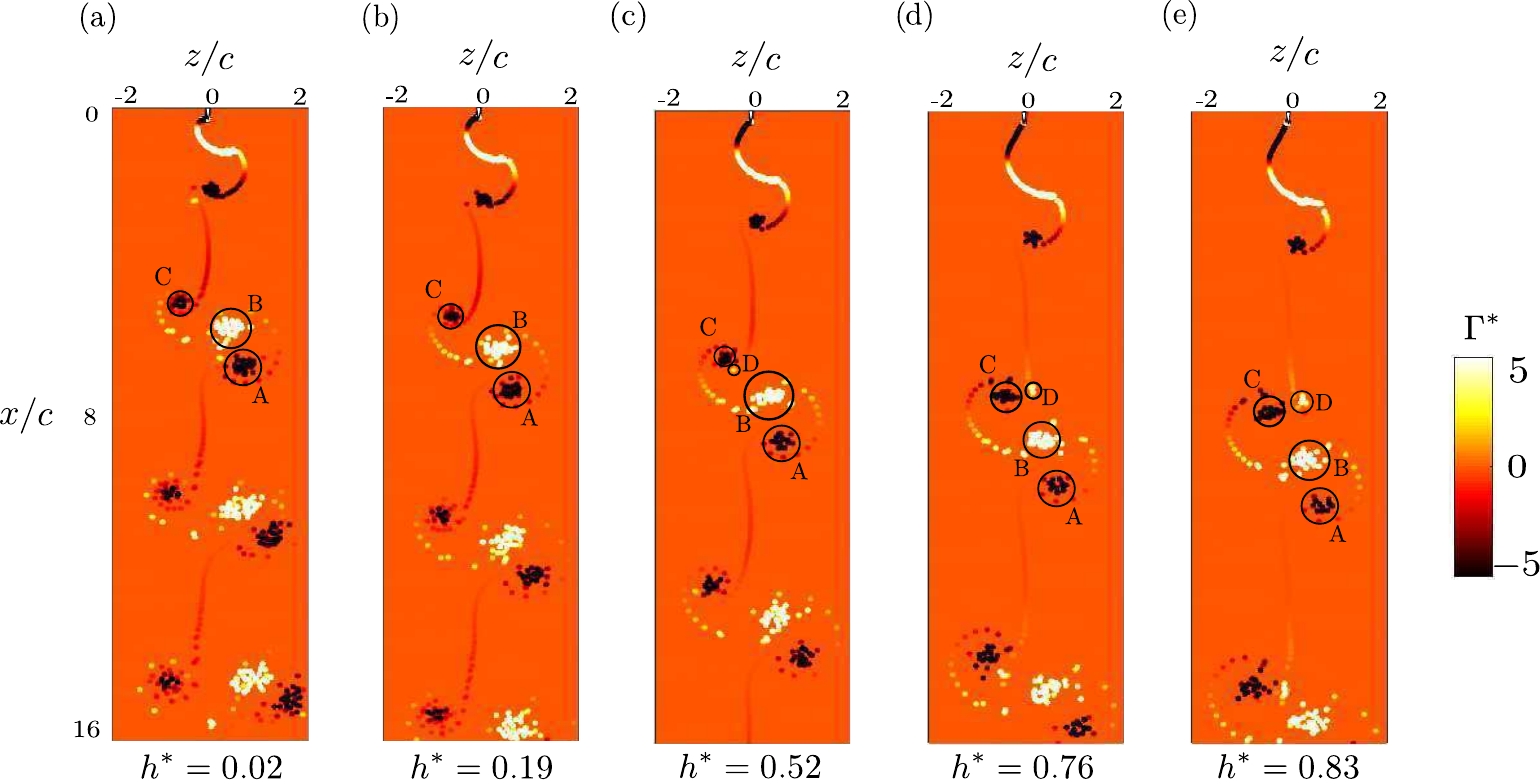}
\caption{\am{The wake flow produced over several oscillation cycles by an intermittent swimmer with  $Li=0.05$, $A^*=0.4$, $DC = 0.5$, and (a) $h^* = 0.02$, (b) $h^* = 0.19$, (c) $h^* = 0.52$, (d) $h^* = 0.76$, and (e) $h^* = 0.83$. The location of the trailing edge of the foil is located at $x/c=0$. The dimensionless circulation of the wake vortices is defined as $\Gamma^* = \Gamma/fAc$.}}\label{flow_field}
\end{figure}

The characteristic wake topology of a purely pitching intermittent swimmer is presented and explained in previous work \cite{Akoz2018}. Here, this understanding is extended to combined heaving and pitching motions.  Figure \ref{flow_field} presents the evolution of the wake flow for a few oscillation cycles of a swimmer at $DC=0.5$ and $A^* = 0.4$. The distribution of positive (anti-clockwise) and negative (clockwise) dimensionless circulation is represented with white and black, respectively. \ea{In Figure \ref{flow_field}a, three vortices are shed per oscillation cycle. As there is no net bound circulation around the hydrofoil at the end of the coasting phase the hydrofoil sheds a starting vortex (vortex A) at the beginning of the motion. Then, two vortices are shed as the hydrofoil changes direction (vortices B and C). However, some combined heaving and pitching motions produce four vortices per bursting cycle. For $h^* > 0.5$ there is a noticeable stopping vortex shed (vortex D) at the end of the bursting phase.  Interestingly, previous work \cite[]{Akoz2018} showed four vortices shed for purely pitching motions ($h^* = 0$), which does not occur here for a lower $A^*$.  This highlights that the strength of the stopping vortex is dependent upon not only the heave ratio, but also the dimensionless amplitude of motion.}

\subsection{Performance Map}

\begin{figure}[t]
\centering
\includegraphics[width=.85\linewidth,scale=0.5]{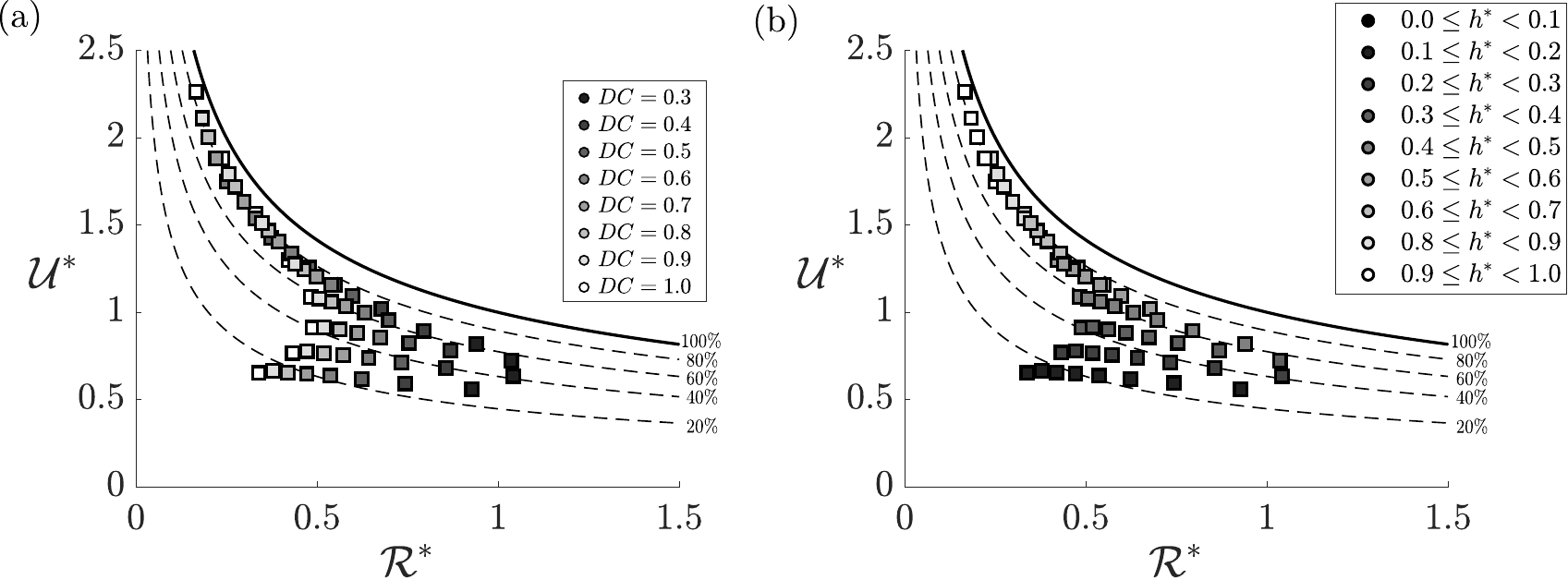}

\caption{\ea{Performance map for $A^* = 1$, $Li = 0.05$. Efficiency of swimmer presented as a function of dimensionless velocity, and dimensionless range, with the color code for variation of (a) duty cycle ($0.3\le DC \le 1$) and (b) heave ratio ($0\le h^* \le 1$). The solid line represents $\eta = 100\%$ and dashed lines represent $\eta = 80 \%, 60 \%, 40 \%,$ and $ 20 \%$.}}\label{performanceMap}
\end{figure}

The relationship among the dimensionless speed, dimensionless range and efficiency of a swimmer can be presented with a performance map as shown in Figure \ref{performanceMap}. Given the definitions of these dimensionless variables in equations (\ref{eq:speed}) and (\ref{eq:range}), efficiency isolines can be defined as,
\begin{align}
  \mathcal{U^*}  = \sqrt{2 \eta/Li} \; \mathcal{R^*}^{-1/2}  
\end{align}

\noindent when the efficiency and Lighthill number are fixed.  Figure \ref{performanceMap} presents the simulation data for \am{$Li = 0.05$} and $A^* = 1$ on the performance map, where the data is a function of $DC$ (Figure \ref{performanceMap}a\am{)} and $h^*$ (Figure \ref{performanceMap}b). The solid line represents the maximum possible efficiency isoline of $\eta=100\%$ while the dashed lines represent efficiency isolines of $\eta =$ 80\%, 60\%, 40\%, and 20\%. Continuous, heave-dominated swimmers can sustain high dimensionless speeds at high efficiency, yet their dimensionless range is limited.  In contrast, intermittent swimmers using moderate $h^*$ values have high dimensionless ranges at relatively low dimensionless speeds. In fact, the lowest duty cycle swimmers have around a two times larger dimensionless range than their continuously swimming counterparts at around $\eta = 45\%$. As dimensionless velocity scales with the thrust coefficient and dimensionless range scales inversely with the power coefficient, the choice of $DC$ and $h^*$ dictate the trade off between maximizing the thrust coefficient and minimizing the power coefficient.

%--------------------------------------------CONCLUSION-------------------------------------------------------%
\section{Conclusions} \label{s:concl}

A self-propelled virtual body connected with a heaving and pitching hydrofoil using continuous or intermittent swimming is examined for a range of dimensionless \ea{heave ratio}s, dimensionless amplitudes, Lighthill numbers, and duty cycles. The maximum angle of attack is restricted to \am{$10^o$} to obtain physically meaningful results with the inviscid simulations. Peak efficiencies around $\eta = 85\%$ occur for continuous swimming using heave-dominated motions ($h^* > 0.7$), which also   generate high dimensionless speeds and low dimensionless ranges. In this heave-dominated regime peak efficiency is increased when $A^*$ increases, and the $Li$ decreases, which corresponds to lower reduced frequencies. Moreover, if the $DC$ is reduced the reduced frequency increases and the efficiency is reduced.

\ea{Switching from a continuous to intermittent gait can be beneficial or detrimental depending upon the continuous swimmer's $h^*$ and $k$.  For pitch-dominated motions, regardless of the continuous swimmer's reduced frequency, switching to intermittent swimming increases the reduced frequency and consequently increase their efficiency.  The largest efficiency benefits of switching to intermittent swimming can be derived for continuous swimmers operating with low reduced frequencies which occur when $Li$ is high and $A^*$ is low for self-propelled motion.  For equally-partitioned heaving and pitching motions, if a continuous swimmer's reduced frequency is low enough then swimming intermittently will increase their reduced frequency and consequently increase their efficiency, however, it could also decrease efficiency if the continuous swimmer's reduced frequency is too high.  For heave-dominated motions, switching to intermittent swimming will increase the swimmer's reduced frequency and result in a decrease in efficiency, which explains the lack of an observed efficiency benefit for $h^* > 0.7$. The qualitative shift in the efficiency trend from heave-dominated to pitch-dominated motions is due to the fundamental nature of the force production. The efficiency of pitch- and heave-dominated motions are maximized by maximizing the added mass based thrust production at high reduced frequencies and the circulatory based thrust production at low reduced frequencies, respectively.}

The wake topology of an intermittent swimmer using a combined heaving and pitching motion is similar to an intermittent swimmer using pure pitching motions. However, for $h^* \leq 0.52$ the stopping vortex shed each cycle becomes negligibly weak and three vortices are shed per oscillation cycle. Intermittent swimming also increases the dimensionless range while maintaining a relatively high propulsive efficiency. Indeed, the lowest duty cycle swimmers have around two times larger dimensionless range than their continuously swimming counterparts.  This work provides a mechanistic understanding of the performance trade-offs that occur between continuous and intermittent motions of combined heaving and pitching hydrofoils, which can provide insight into biological locomotion.  For example, in nature intermittent swimmers such as trout \cite{Yanase2015,Yanase2016}, koi carps \cite{Wu2007}, and cod \cite{Videler1981} tend to use pitch-dominated motions of their caudal fins, while continuous swimmers such as cetaceans \cite{fish93b,fish1998comparative,fish1998biomechanical} tend to use heave-dominated motions.  Now, we can hypothesize that biological pitch-dominated swimmers may enjoy enhanced efficiency by using intermittent motions, while biological heave-dominated swimmers have nothing to gain by using intermittent motions.

\section{Acknowledgements}
This work was supported by the Office of Naval Research under Program Director Dr. Robert Brizzolara on MURI grant number N00014-08-1-0642.

\bibliography{MAIN}

\section{Appendix}
\subsection{Convergence}\label{App conv}
\begin{figure}[h]
    \centering
    \includegraphics[width=0.85\linewidth, scale=0.75]{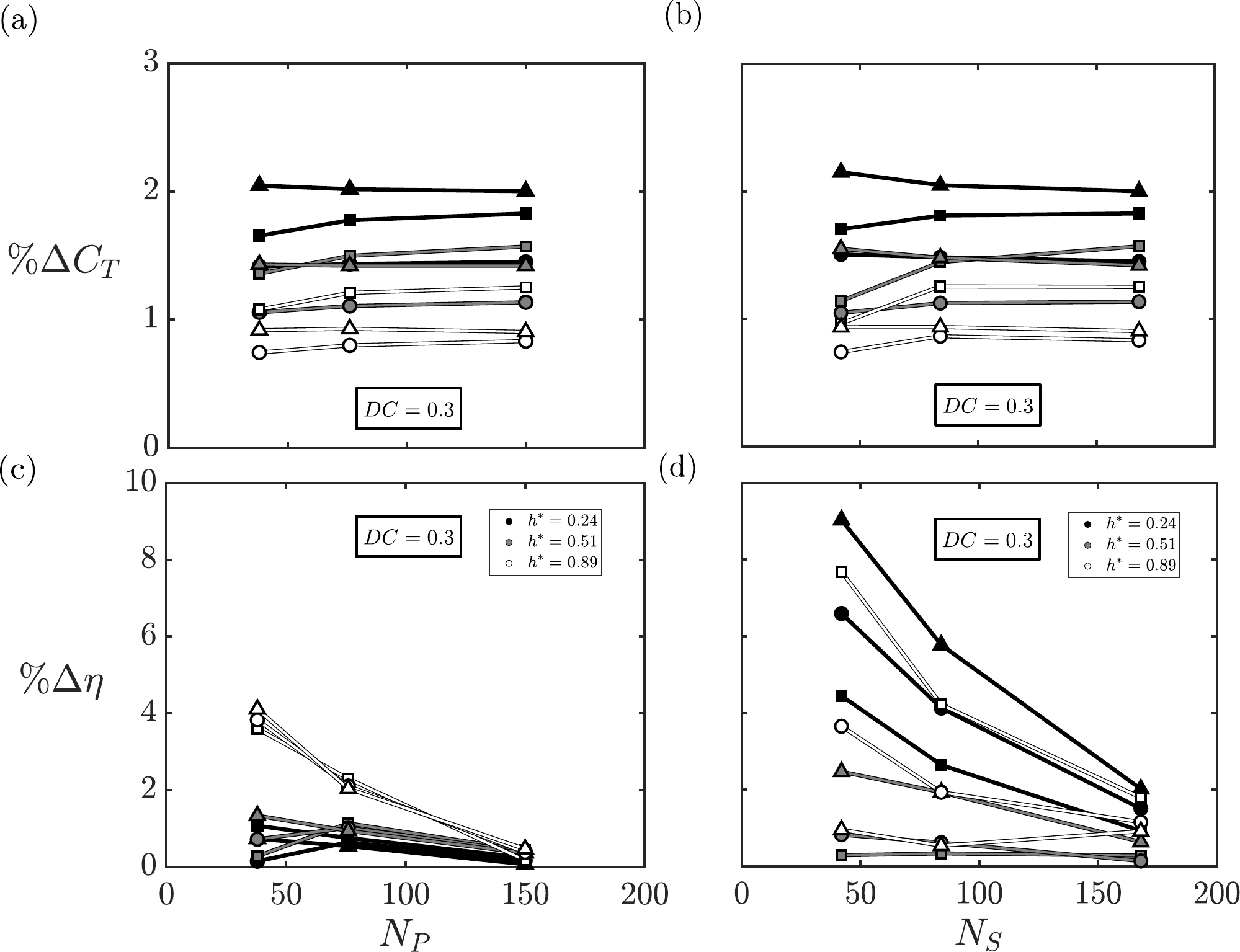}
    \caption{Convergence plots for $DC=0.3$ and $A^*=1$}
    \label{fig:convergeDC3}
\end{figure}

\begin{figure}[h]
    \centering
    \includegraphics[width=0.85\linewidth, scale=0.75]{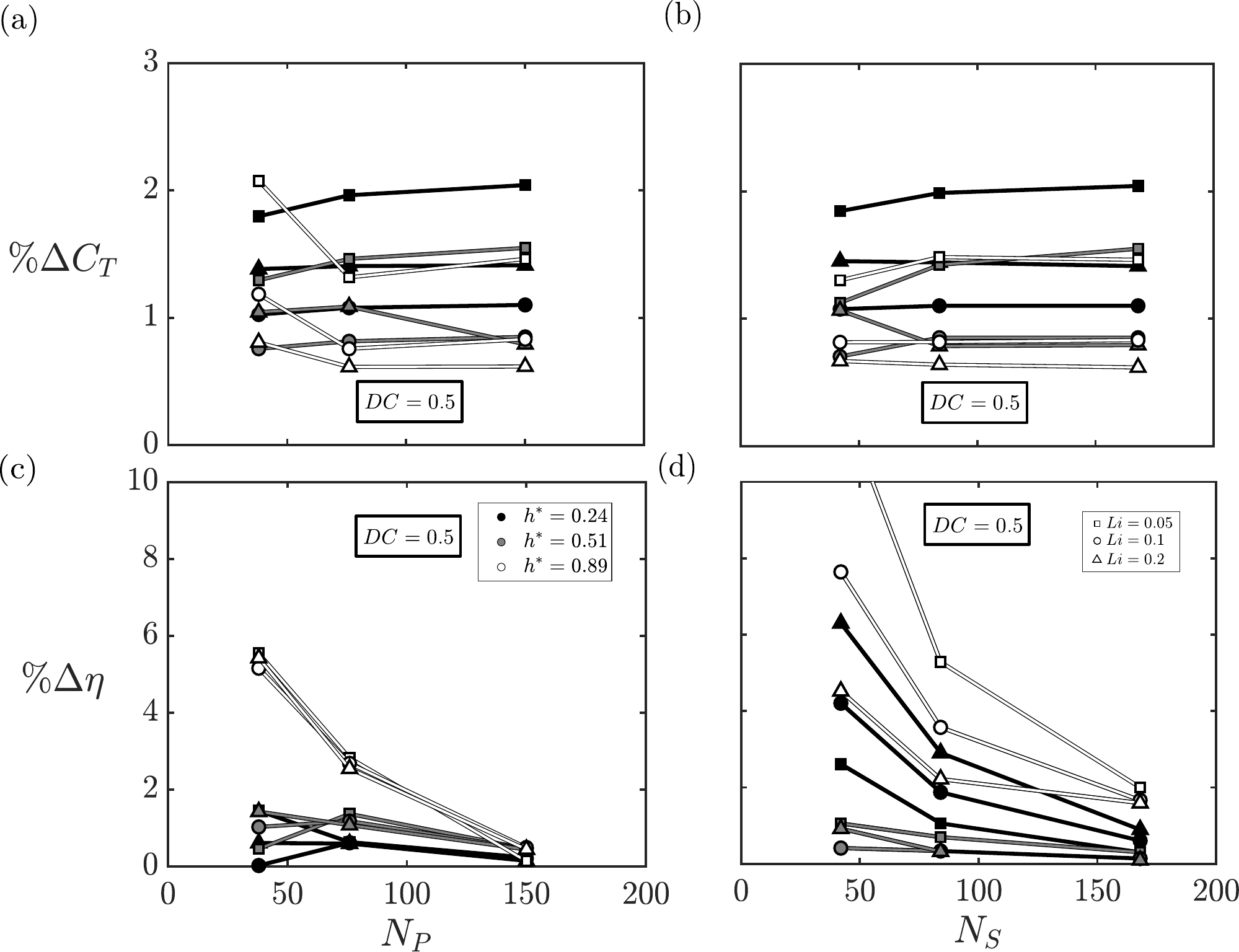}
    \caption{Convergence plots for $DC=0.5$ and $A^*=1$}
    \label{fig:convergeDC5}
\end{figure}

\begin{figure}[h]
    \centering
    \includegraphics[width=0.85\linewidth, scale=0.75]{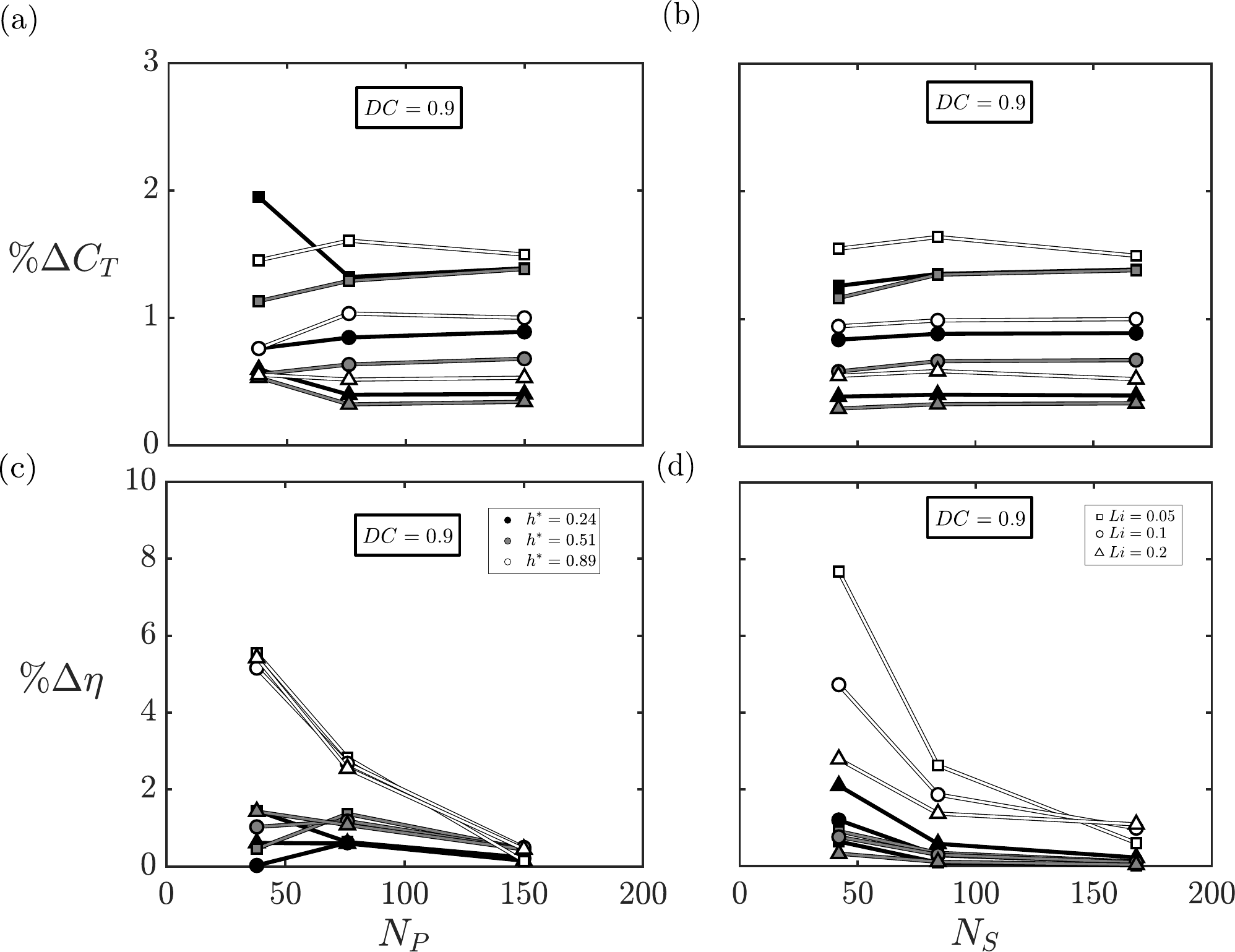}
    \caption{Convergence plots for $DC=0.9$ and $A^*=1$}
    \label{fig:convergeDC9}
\end{figure}

\ea{Discretization independence graphs for a maximum dimensionless amplitude of $A^*=1$, three different duty cycles of $DC=0.3, 0.5$ and $0.9$, and three different heave ratios of $h^*= 0.24, 0.51$ and $0.89$ are shown in Figures \ref{fig:convergeDC3}-\ref{fig:convergeDC9}. In all of the mentioned figures, the percent change in (a) the thrust coefficient and (c) efficiency are presented when the number of time steps are held constant ($N_S=168$) and the number of body elements are doubled. Similarly, discretization independence graphs showing the percent change in (b) the thrust coefficient and (d) efficiency are presented when the number of body elements are held constant ($N_P=150$) and the number of time steps per bursting period are doubled.  The thrust and efficiency change by less than approximately 2\% when the number of body elements and time steps per bursting period are doubled from $N_P=150$ and $N_S=168$, respectively.}

\end{document}